\documentclass[twocolumn]{aastex62}

\usepackage{graphicx}
\usepackage{color}
\usepackage{mathrsfs,amsmath,amssymb}
\usepackage{ulem}
\usepackage{url}

\newcommand{\hb}{\ifmmode {\rm H\beta} \else H$\beta$\fi}
\newcommand{\mgii}{\ifmmode {\rm Mg\ II} \else Mg {\sc ii}\fi}
\newcommand{\feii}{\ifmmode {\rm Fe\ II} \else Fe {\sc ii}\fi}
\newcommand{\heii}{\ifmmode {\rm He\ II} \else He {\sc ii}\fi}
\newcommand{\oiii}{\ifmmode {\rm [O\ III]} \else [O {\sc iii}]\fi}
\newcommand{\mbh}{\ifmmode {M_{\bullet}} \else $M_{\bullet}$\fi}
\newcommand{\fblr}{\ifmmode {f_{\rm BLR}} \else $f_{\rm BLR}$\fi}
\newcommand{\rhb}{\ifmmode {R_{\rm H\beta}} \else $R_{\rm H\beta}$\fi}
\newcommand{\rblr}{\ifmmode {R_{\rm BLR}} \else $R_{\rm BLR}$\fi}
\newcommand{\taublr}{\ifmmode {\tau_{\rm BLR}} \else $\tau_{\rm BLR}$\fi}
\newcommand{\tauhb}{\ifmmode {\tau_{\rm H\beta}} \else $\tau_{\rm H\beta}$\fi}

\graphicspath{{./}{figures/}}

\begin{document}

\title{MONITORING AGNS WITH \hb\ ASYMMETRY. II.  \\
Reverberation Mapping of Three Seyfert Galaxies Historically Displaying \hb\ Profiles with Changing Asymmetry: \\
Mrk 79, NGC 3227, and Mrk 841}

\correspondingauthor{Michael Brotherton}
\email{mbrother@uwyo.edu}

\author{Michael S. Brotherton}
\affiliation{Department of Physics and Astronomy, University of Wyoming, Laramie, WY 82071, USA; \url{mbrother@uwyo.edu}}

\author{Pu Du}
\affiliation{Key Laboratory for Particle Astrophysics, Institute of High Energy Physics,
Chinese Academy of Sciences, 19B Yuquan Road, Beijing 100049, China;\\ }

\author{Ming Xiao}
\affiliation{Key Laboratory for Particle Astrophysics, Institute of High Energy Physics,
Chinese Academy of Sciences, 19B Yuquan Road, Beijing 100049, China;\\ }

\author{Dong-Wei Bao}
\affiliation{Key Laboratory for Particle Astrophysics, Institute of High Energy Physics,
Chinese Academy of Sciences, 19B Yuquan Road, Beijing 100049, China;\\ }

\author{Bixuan Zhao}
\affiliation{Physics Department, Nanjing Normal University, Nanjing 210097, China}

\author{Jacob N. McLane}
\affiliation{Department of Physics and Astronomy, University of Wyoming, Laramie, WY 82071, USA; \url{mbrother@uwyo.edu}}

\author{Kianna A. Olson}
\affiliation{Department of Physics and Astronomy, University of Wyoming, Laramie, WY 82071, USA; \url{mbrother@uwyo.edu}}

\author{Kai Wang}
\affiliation{Key Laboratory for Particle Astrophysics, Institute of High Energy Physics,
Chinese Academy of Sciences, 19B Yuquan Road, Beijing 100049, China;\\ }

\author{Zheng-Peng Huang}
\affiliation{Key Laboratory for Particle Astrophysics, Institute of High Energy Physics,
Chinese Academy of Sciences, 19B Yuquan Road, Beijing 100049, China;\\ }

\author{Chen Hu}
\affiliation{Key Laboratory for Particle Astrophysics, Institute of High Energy Physics,
Chinese Academy of Sciences, 19B Yuquan Road, Beijing 100049, China;\\ }

\author{David H. Kasper}
\affiliation{Department of Physics and Astronomy, University of Wyoming, Laramie, WY 82071, USA; \url{mbrother@uwyo.edu}}

\author{William T. Chick}
\affiliation{Department of Physics and Astronomy, University of Wyoming, Laramie, WY 82071, USA; \url{mbrother@uwyo.edu}}

\author{My L. Nguyen}
\affiliation{Department of Physics and Astronomy, University of Wyoming, Laramie, WY 82071, USA; \url{mbrother@uwyo.edu}}

\author{Jaya Maithil}
\affiliation{Department of Physics and Astronomy, University of Wyoming, Laramie, WY 82071, USA; \url{mbrother@uwyo.edu}}

\author{Derek Hand}
\affiliation{Department of Physics and Astronomy, University of Wyoming, Laramie, WY 82071, USA; \url{mbrother@uwyo.edu}}

\author{Yan-Rong Li}
\affiliation{Key Laboratory for Particle Astrophysics, Institute of High Energy Physics,
Chinese Academy of Sciences, 19B Yuquan Road, Beijing 100049, China;\\ %\url{dupu@ihep.ac.cn}, \url{wangjm@ihep.ac.cn}
}

\author{Luis C. Ho}
\affiliation{Kavli Institute for Astronomy and Astrophysics, Peking University, Beijing 100871, China}
\affiliation{Department of Astronomy, School of Physics, Peking University, Beijing 100871, China}

\author{Jin-Ming Bai}
\affiliation{Yunnan Observatories, Chinese Academy of Sciences, Kunming 650011, China}

\author{Wei-Hao Bian}
\affiliation{Physics Department, Nanjing Normal University, Nanjing 210097, China}

\author{Jian-Min Wang}
\altaffiliation{PI of the MAHA Project}
\affiliation{Key Laboratory for Particle Astrophysics, Institute of High Energy Physics,
Chinese Academy of Sciences, 19B Yuquan Road, Beijing 100049, China;\\ %\url{dupu@ihep.ac.cn}, \url{wangjm@ihep.ac.cn}
}
\affiliation{National Astronomical Observatories of China, Chinese Academy of Sciences, 20A Datun Road, Beijing 100020, China}
\affiliation{School of Astronomy and Space Science, University of Chinese Academy of Sciences, 19A Yuquan Road, Beijing 100049, China}

\collaboration{(MAHA Collaboration)}

\begin{abstract} 
We report the results of reverberation mapping three bright Seyfert galaxies, Mrk 79, NGC 3227, and Mrk 841, from a campaign conducted from December 2016 to May 2017 with the Wyoming Infrared Observatory (WIRO) 2.3-meter telescope.
All three of these targets have shown asymmetric broad H$\beta$ emission lines in the past, although their emission lines were relatively symmetric during our observations.
We measured \hb\ time lags for all three targets and estimated masses of their black holes -- for the first time in the case of Mrk 841.  
For Mrk 79 and NGC 3227, the data are of sufficient quality to resolve distinct time lags as a function of velocity and to compute two-dimensional velocity-delay maps.
Mrk 79 shows smaller time lags for high-velocity gas but the distribution is not symmetric, and its 
complex velocity-delay map could result from the combination of both inflowing and outflowing \hb\ emitting disks that may be part of a single 
larger structure.
NGC~3227 shows the largest time lags for blueshifted gas and the two-dimensional velocity-delay map suggests a disk with some inflow.
We compare our results with previous work and find evidence for different time lags despite similar luminosities, as well as evolving broad line region structures.

\end{abstract}


\section{Introduction}

Reverberation mapping (RM) is a technique based on the spectroscopic monitoring of Seyfert galaxies and quasars \citep[e.g.,][]{bahcall1972, blandford1982, peterson1993}.  
The time delay between the variable continuum and a corresponding response from a broad emission line (usually \hb\ in practice), permits the measurement of the emissivity-weighted size of the corresponding broad line region (BLR), since the speed of light is finite
and the recombination timescale short.
In combination with a velocity, obtained from the Doppler-broadened
emission-line profile (usually the variable part as characterized by the root-mean-square profile in practice), the time lag can be used to estimate a virial mass of the central
supermassive black hole  \citep[e.g.,][]{peterson2004}.

The black hole masses of active galactic nuclei (AGNs) are arguably their most fundamental property. Despite great interest in the determination of black hole masses and decades of efforts, the substantial amount of observational resources required for RM has limited good time lag measurements to on the order of 100 objects \citep[e.g.,][]{kaspi2000,peterson2004,bentz2015, barth2015,fausnaugh2017,grier2017,du2018a,derosa2018}.  

With sufficiently high cadence, long-duration campaigns, sufficient signal-to-noise ratios (SNRs), and spectral resolution, more detailed analyses of RM data  can yield substantially more information about the BLR.  Specifically, RM campaigns have begun to more regularly produce 
velocity-resolved time lags, for instance, 
\citep{kollatschny2003,bentz2008,bentz2009,denney2009,denney2010,barth2011,doroshenko2012,grier2013,valenti2015,barth2015,skielboe2015,lu2016,du2016VI,du2018b,derosa2018}.
These efforts provide evidence that the BLR has a range of kinematics present in different objects, with a mix of flattened disks, infall, and perhaps less often, outflow.  
A few AGNs have also shown more complex patterns that do not fit neatly into these simple categories \citep[e.g.,][]{du2018b,derosa2018}.  There does not seem to be a single common BLR structure
with uniform kinematics.

The best quality ``high-fidelity'' RM data sets can be used to produce two-dimensional velocity-delay maps (VDMs) \citep[e.g.,][]{horne2004}.  A VDM is a model that separates the average emission-line response for a given velocity into a distribution of time lags, and in principle can account for some of the degeneracy inherent in line-of-sight velocities for a three-dimensional object.  A VDM has a second dimension compared to velocity-resolved time lags, and is therefore in principle can be much more diagnostic of the properties 
of the BLR.  VDMs have been produced for a couple of dozen objects \citep[e.g.,][]{grier2013,kollatschny2014,skielboe2015,xiao2018a,xiao2018b}.  

Recent RM campaigns discovered that the geometry or kinematics of BLRs can change dramatically over relatively short periods, e.g., the BLR of NGC 3227 transformed from outflow in 2007 to virialized motion in 2012, while the BLR of NGC 3516 went from infall in 2007 to outflow in 2012 \cite[see][]{denney2010, derosa2018}. These phenomena reveal that the BLRs of AGNs are evolving and changing rather than static in a timescale of only a few years. 
Furthermore, the BLR of NGC 5548 shows both long-term evolution \citep{xiao2018b} and short-term abnormal behavior \citep{pei2017}.
The physical reason for these changes is far from being fully understood, which makes it necessary to monitor the same targets in different years.

We recently initiated an RM project using the Wyoming Infrared Observatory (WIRO) 2.3 meter telescope.  The project is titled ``Monitoring Active Galactic Nuclei with \hb\ Asymmetry,'' or MAHA, and has the goal of obtaining high-quality, high-cadence RM data, primarily of AGNs with asymmetric profiles. Investigating asymmetric profiles with RM will permit the exploration of the full diversity of the BLR as well as searching for AGNs powered by binary black holes.  No simple relationship between the profiles of AGN emission lines and the kinematical structure of their BLRs has yet been established, although some correspondence might be naively expected. Furthermore, the objects with asymmetric \hb\ at present or historically must have extreme kinematics (e.g., fast outflow or inflow) or significant variability, thus, are the ideal targets for the investigation of BLR evolution. We recently reported first results from the project for four objects with asymmetric \hb\ lines \citep{du2018b}.

In our initial campaign, we also observed several Seyfert galaxies, Mrk 841, Mrk 79, and NGC 3227, that had displayed very asymmetric \hb\ lines in the past, but appeared more symmetric in 2017. In the early 1980s, Mrk 841 (also known as PG 1501+106) had a red asymmetric \hb\ line (showing a significant tail of emission extending on the red side of the profile compared to the blue side), while Mrk 79 and NGC 3227 had rather symmetric profiles then \citep{DeRobertis1985}.  Through at least the mid-1990s, Mrk 841 still showed a red asymmetry in \hb\ \citep{boroson1992,jansen2000}. Mrk 79 developed a strong blue asymmetry in \hb\ during the 1990s \citep{peterson1998}.  During 2007, NGC 3227 possessed an \hb\ profile with a strong red asymmetry and a double-peaked profile in its rms spectrum \citep{denney2009}.  In 2011, Mrk 841's \hb\ line showed a {\em blue} asymmetry \citep{barth2015}, very different from its red asymmetry in the 1980s and 1990s.  We monitored these Seyfert galaxies with the prospect of seeing a transition to a more asymmetric \hb\ profile, and their corresponding BLR kinematics before, after, and perhaps during such a transition.

We report our results for our initial RM campaign of Mrk~79, NGC~3227, and Mrk~841.  We focus on the H$\beta$ line, and particularly on profile-dependent variations that can reveal BLR kinematics. Section 2 describes our observations and data reduction, as well as the construction of our continuum and \hb\ light curves.  Section 3 describes our analyses, which include the determination of the integrated \hb\ line time lags, the calculation of the corresponding black hole masses, and the measurement of the velocity-resolved time lags. For Mrk 79 and NGC 3227 we were also able to determine VDMs, which we present in section 4. In section 5 we discuss our results for each object, comparing the new results to past results as appropriate.  Section 6 describes our plans for future work and 7 summarizes our conclusions.

\section{Observations and Data Reduction}
\label{sec:methods}

\subsection{Observations}

We conducted our observations between late December 2016 and late May 2017.
Table \ref{tab:obj} provides the positions and redshifts of our targets, along with the number of spectra and average cadence. Below we describe our observations and data reduction.

%----------------------------- Table 1 -------------------------------

\begin{deluxetable}{lllccc}
%\rotate
\tablecolumns{6}
\tablewidth{\textwidth}
\setlength{\tabcolsep}{4.5pt}
\tablecaption{Targets\label{tab:obj}}
\tabletypesize{\footnotesize}
\tablehead{
\colhead{Object}                      &
\colhead{$\alpha_{2000}$}             &
\colhead{$\delta_{2000}$}             &
\colhead{Redshift}                    &
\colhead{$N_{\rm spec}$}              &
\colhead{Cadence}                                     
}
\startdata
Mrk~79       & 07 42 32.8 & +49 48 35 & 0.0222 & 40 & 3.8  \\
NGC~3227     & 10 23 30.6 & +19 51 54 & 0.0039 & 48 & 3.1  \\
Mrk~841      & 15 04 01.2 & +10 26 16 & 0.0364 & 17 & 6.4  
\enddata
\tablecomments{$N_{\rm spec}$ is the number of spectroscopic epochs. 
``Cadence'' is the average sampling interval in days
in the observed frame.
}
\end{deluxetable}

\subsubsection{Spectrophotometry}

We used WIRO and its Long Slit Spectrograph in remote-operation mode   \citep{findlay2016}.
The 900 line mm$^{-1}$ grating provided a dispersion of 1.49 \AA\ pixel$^{-1}$ while covering the range $\sim$4000 -- 7000 \AA.  We used a 5$^{\prime\prime}$-wide slit oriented north-south. We typically took three exposures of 5 minutes each but would take more exposures during poor observing conditions.  We observed at least one spectrophotometric standard star each night, usually Feige 34 but also G1912B2.  We observed CuAr lamps for wavelength calibration and a uniformly illuminated white card inside the dome for our flat fields.

We employed standard data reduction techniques with IRAF v2.16.
We extracted our spectra using apertures of $\pm6^{\prime\prime}.8$
and background windows of $7^{\prime\prime}.6$ -- $15^{\prime\prime}.1$ on both sides.  We performed flux calibrations and extinction corrections using standard 
techniques, but clouds and potential slit losses required additional
calibrations to obtain spectrophotometric data products.

The \oiii\ $\lambda\lambda$ 4959, 5007 lines come from an extended narrow-line region, which only varies very slowly over many years, allowing them to be used for flux calibration within the duration of an observing campaign of months \citep{peterson2013}.
We employed the calibration technique of \cite{vanGroningen1992},
with minor modifications,
as described by \citet{du2018b}.
Our [O III]-based procedures 
provide relatively high-precision light curves.

We measured the \oiii\ fluxes using the wavelength windows listed in Table
\ref{tab:windows}. We determined the \oiii\ fluxes (also given in Table
\ref{tab:windows}) using spectra taken in photometric conditions. 
We obtained final flux-calibrated spectra by averaging
(appropriately noise-weighted) exposures in the same night for each object.

\begin{figure}
\centering
\includegraphics[width=0.45\textwidth]{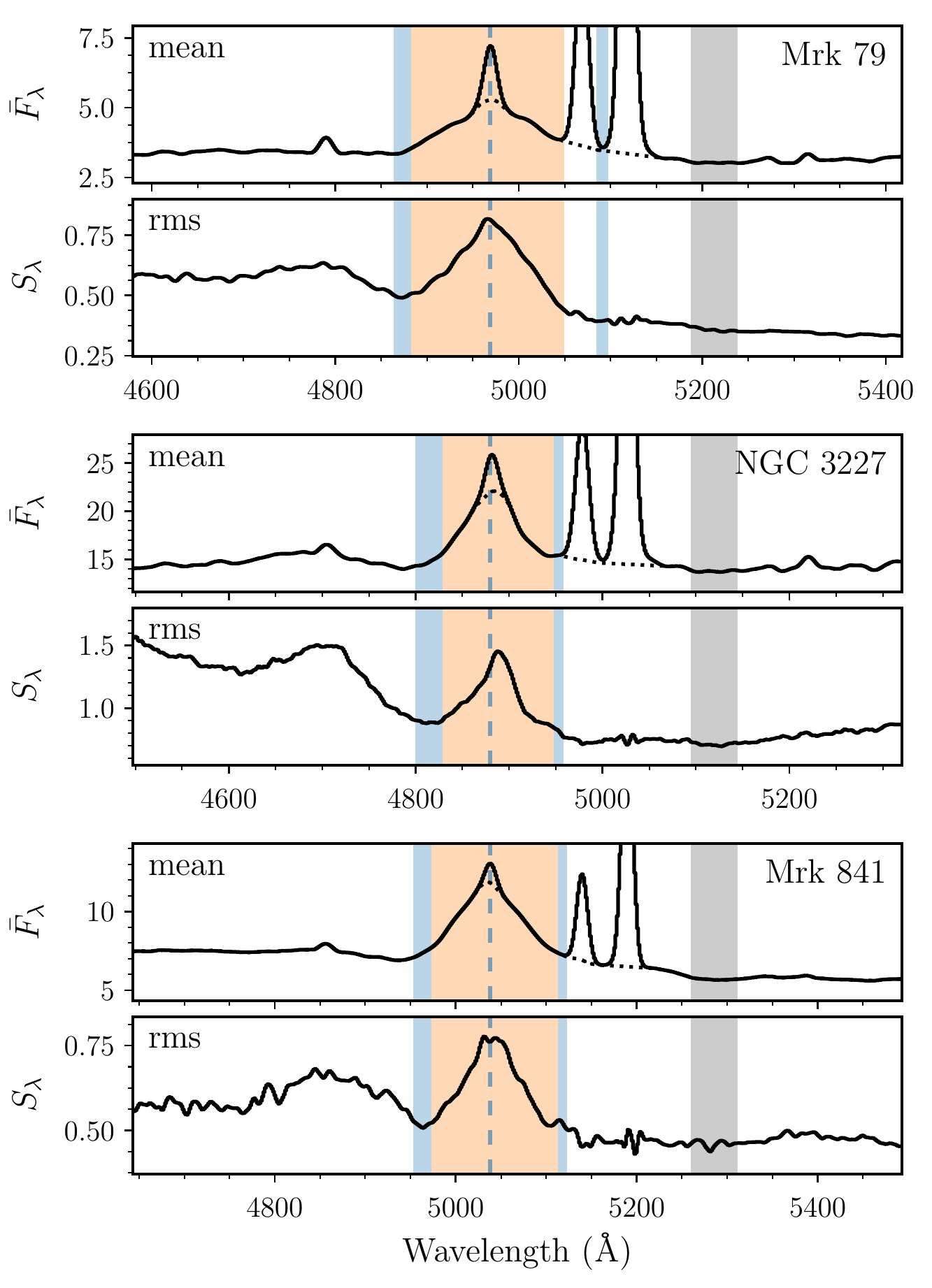}
\caption{Mean and rms spectra.  The orange and blue regions show the windows for \hb\ emission lines and their backgrounds. The gray regions indicate the 5100\AA\ continuum windows. We plot $4861\times(1+z)$ \AA\ as vertical dashed lines in order to illustrate zero velocity clearly. Dotted lines show the our narrow line fitting in the mean spectra.The flux density units are 
$10^{-15}$ erg s$^{-1}$ cm$^{-2}$ \AA$^{-1}$.}
\label{fig:meanrms}
\end{figure}

We calculated mean and root-mean-square (rms) spectra (Figure \ref{fig:meanrms}). The narrow  \oiii\ lines have essentially vanished in the rms spectra, indicating that our flux calibration procedure works well.  Some examples of the flux-calibrated spectra are shown in Appendix \ref{app:spectra}.

The rms spectrum of NGC 3227 slightly increases to the red, a little unexpected, 
apparently due to  varying host contamination in the slit caused by tracking
inaccuracy. The host galaxy of NGC 3227 is the strongest relative to the AGN
in the present sample, and during five of the last nine epochs there was more significant host contamination relative to the other points,
which makes the slopes of these spectra steeper toward redder wavelengths. The relative
increase of the host contamination of these five epochs also makes their continuum fluxes
shift upward a little in the following Figure \ref{fig:light_curves}, but the
shift amplitudes are within $\sim5\%$ compared with the adjacent points. The results for NGC 3227 in the following time-lag analysis are
consistent with/without these five epochs. Therefore, we still keep them for completeness.

%Table 2

\begin{deluxetable*}{lccccccccc}
%\rotate
\tablecolumns{10}
\tablewidth{\textwidth}
\setlength{\tabcolsep}{4pt}
\tablecaption{Measurement windows in the observed-frame and \oiii\ standard fluxes\label{tab:windows}}
\tabletypesize{\footnotesize}
\tablehead{
\colhead{}                            &
\colhead{}                            &
\colhead{}                            &
\multicolumn{3}{c}{\oiii}             &
\colhead{}                            &
\multicolumn{3}{c}{\hb}               \\ \cline{4-6} \cline{8-10}
\colhead{Object}                      &
\colhead{$F_{\oiii}$}                 &
\colhead{}                            &
\colhead{continuum (blue)}            &
\colhead{line}                        &
\colhead{continuum (red)}             & 
\colhead{}                            &  
\colhead{continuum (blue)}            &
\colhead{line}                        &
\colhead{continuum (red)}             \\ 
\colhead{}                            &
\colhead{($10^{-13}\ \mathrm{erg\ s^{-1}\ cm^{-2}}$)} &
\colhead{}                            &
\colhead{(\AA)}                       &
\colhead{(\AA)}                       &
\colhead{(\AA)}                       &
\colhead{}                            &
\colhead{(\AA)}                       &
\colhead{(\AA)}                       &
\colhead{(\AA)}                            
}
\startdata
Mrk~79       & 3.61 & & 5085--5095 & 5095--5145 & 5145--5165 & & 4864--4883 & 4883--5050 & 5084--5098  \\
NGC~3227     & 7.70 & & 4995--5005 & 5005--5050 & 5050--5070 & & 4799--4829 & 4829--4948 & 4948--4958  \\
Mrk~841      & 2.52 & & 5155--5170 & 5170--5215 & 5215--5235 & & 4953--4973 & 4973--5113 & 5113--5123  
\enddata
%\tablecomments{
%}
\end{deluxetable*}

\subsection{WIRO Light Curves}

Figure \ref{fig:light_curves} and 
Table \ref{tab:light_curves} provide the 5100 \AA\ continuum and \hb\ emission-line light curves.
We measure continuum and \hb\ emission-line fluxes from our calibrated spectra. Table \ref{tab:windows} provides the wavelengths of our windows for measurements, while Figure \ref{fig:meanrms} shows them graphically. We measured the 5100 \AA\ continuum values by averaging the fluxes in the window from 5075 to 5125 \AA\ in the rest frame (gray regions in Figure \ref{fig:meanrms}).

We chose continuum background windows for the \hb\ flux measurements 
(light blue regions in Figure \ref{fig:meanrms}) that avoid the \oiii\ lines.  We chose integration windows that cover the \hb\ emission in the rms spectra (orange regions in Figure \ref{fig:meanrms}).  A linear interpolation is used to determine the
continuum to subtract in order to obtain the integrated \hb\ line fluxes.

The flux uncertainties in the resulting light curves include two contributions: 
(1) Poisson noise and (2) systematic uncertainties. The systematic uncertainties result primarily from tracking errors and variable 
observing conditions (e.g., variable extinction, seeing, etc.); we estimated these using the scatter in the flux of consecutive exposures over $\sim$4700-5100 \AA.
We add the two contributions in quadrature and plot them as the error bars in Figure \ref{fig:light_curves}.  This procedure likely underestimates the true uncertainties, as flux differences between adjacent nights are sometimes larger than the error bars. We provide additional systematic uncertainties estimated when significant using 
the median filter method \citep{du2014, du2015} and show them as the gray error bars in the lower-right corners of Figure \ref{fig:light_curves}. All these uncertainties are taken into account in our subsequent time-series analyses.

%--------------------------------- Table 3 ---------------------------------
\begin{deluxetable*}{ccccccccccc}
%\rotate
\tablecolumns{11}
\tablewidth{\textwidth}
\setlength{\tabcolsep}{5pt}
\tablecaption{Light curves\label{tab:light_curves}}
\tabletypesize{\footnotesize}
\tablehead{
\multicolumn{3}{c}{Mrk~79}       &
\colhead{}                       &
\multicolumn{3}{c}{NGC~3227}      &
\colhead{}                       &
\multicolumn{3}{c}{Mrk~841}        \\ \cline{1-3} \cline{5-7} \cline{9-11}
\colhead{JD}                     &
\colhead{$F_{5100}$}             &
\colhead{$F_{\rm H\beta}$}       &
\colhead{}                       &  
\colhead{JD}                     &
\colhead{$F_{5100}$}             &
\colhead{$F_{\rm H\beta}$}       &
\colhead{}                       &
\colhead{JD}                     &
\colhead{$F_{5100}$}             &
\colhead{$F_{\rm H\beta}$}                        
}
\startdata
   52.72 & $  3.71\pm  0.03$ & $  1.93\pm  0.02$ & &    46.87 & $ 12.39\pm  0.02$ & $  4.05\pm  0.02$ & &    98.96 & $  5.33\pm  0.09$ & $  3.40\pm  0.06$\\ 
   71.69 & $  3.29\pm  0.01$ & $  2.17\pm  0.01$ & &    52.79 & $ 13.32\pm  0.01$ & $  4.20\pm  0.02$ & &   100.97 & $  5.11\pm  0.03$ & $  3.41\pm  0.02$\\ 
   72.65 & $  3.37\pm  0.01$ & $  2.14\pm  0.01$ & &    54.90 & $ 13.96\pm  0.01$ & $  4.36\pm  0.02$ & &   103.01 & $  5.07\pm  0.05$ & $  3.25\pm  0.04$\\ 
   74.73 & $  3.36\pm  0.02$ & $  2.12\pm  0.01$ & &    71.87 & $ 14.75\pm  0.06$ & $  5.36\pm  0.03$ & &   105.98 & $  5.09\pm  0.02$ & $  3.35\pm  0.02$\\ 
   79.91 & $  3.56\pm  0.03$ & $  2.22\pm  0.02$ & &    72.89 & $ 14.71\pm  0.03$ & $  5.58\pm  0.02$ & &   115.95 & $  5.11\pm  0.01$ & $  3.30\pm  0.01$
\enddata
  \tablecomments{JD: Julian dates from 2,457,700; $F_{5100}$ and $F_{\hb}$ are the continuum fluxes at 5100 \AA\ and
      \hb\ fluxes in units of $10^{-15}\ {\rm erg\ s^{-1}\ cm^{-2}\ \AA^{-1}}$ and
      $10^{-13}\ {\rm erg\ s^{-1}\ cm^{-2}}$, respectively. This table is available in its entirety in a machine-readable form in the online journal. A portion is shown here for guidance regarding its form and contents.
      }
\end{deluxetable*}

\subsection{Continuum Light Curves from ASAS-SN Photometry}

Given that our targets are relatively bright, we follow \citet{du2018b} in supplementing our continuum light curves with available survey data. Because of the relatively weak emission-line contribution to the total flux in a broadband filter, photometry can be used to supplement continuum light curves, yielding potential improvements in temporal coverage and cadence. The All-Sky Automated Survey for Supernovae  (ASAS-SN)\footnote{\url{http://www.astronomy.ohio-state.edu/~assassin/index.shtml}} provides photometric light curves for relatively isolated objects down to $\sim$17 magnitude   \citep{shappee2014,kochanek2017},
including our targets during our campaign.  This was especially useful for Mrk~841.

Some care scaling the ASAS-SN photometry  is necessary, however, given that its aperture has a radius of 15$^{\prime\prime}$.6),
different from that of our spectroscopy.  This will mean contamination from extra host galaxy light for these low-redshift targets.

We assumed that
\begin{equation}
\label{eqn:scale}
F_{5100} = A + B \times F_{\rm ASAS-SN}, 
\end{equation}
where $A$ is a flux adjustment, and $B$ is a scale factor.  Using ASAS-SN photometry within two days of $F_{5100}$ measurements, we empirically determined $A$ and $B$ for each object by 
fitting our data using the FITEXY algorithm \citep{press1992}.
We removed some extreme ASAS-SN outliers with large error bars. We combined the adjusted and scaled ASAS-SN data with our $F_{5100}$ fluxes, averaging data from the same nights. Figure \ref{fig:light_curves} shows both the adjusted and scaled ASAS-SN light curves as well as the combined light curves.

\begin{figure*}
\centering
\includegraphics[width=0.75\textwidth]{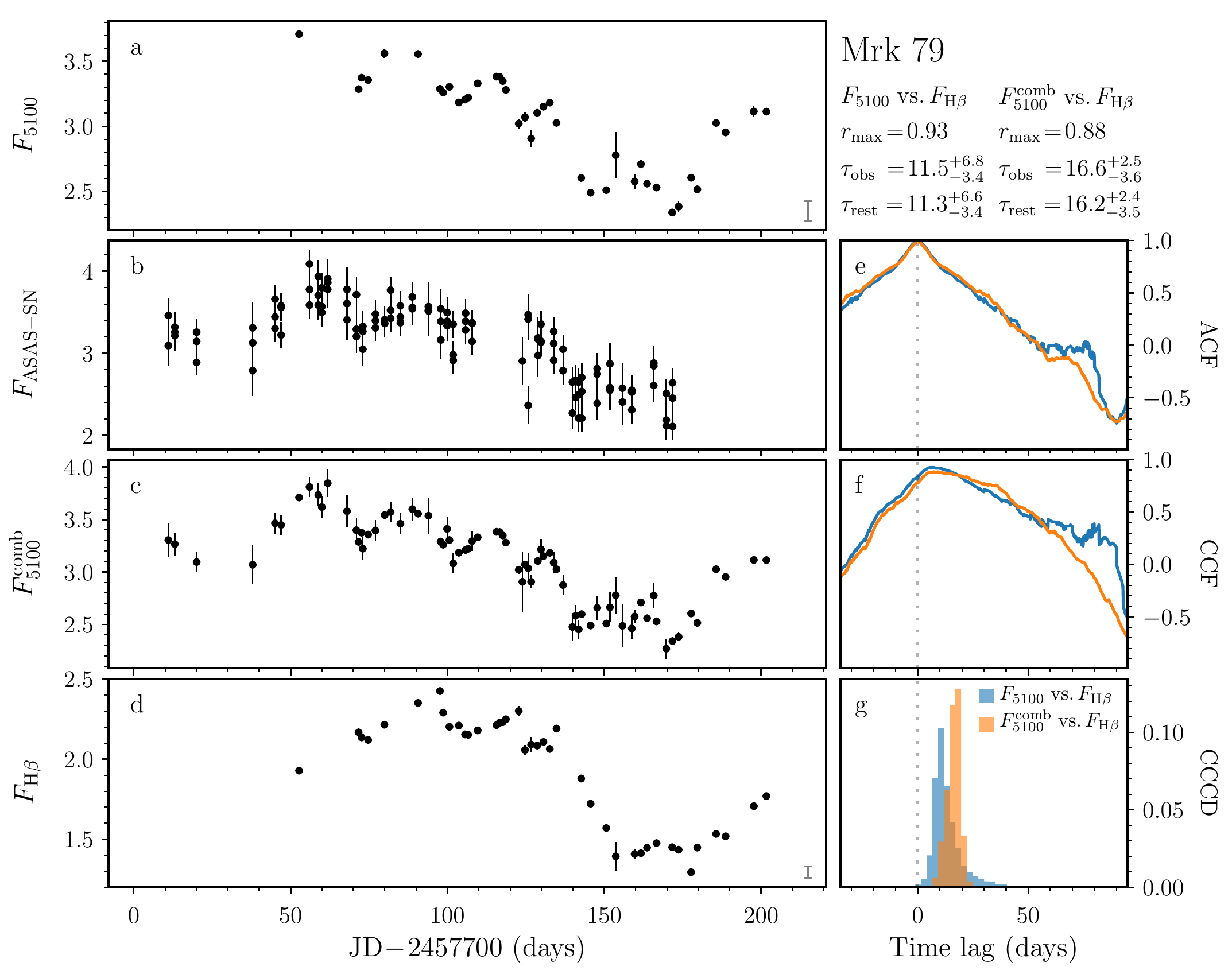} \\
\includegraphics[width=0.75\textwidth]{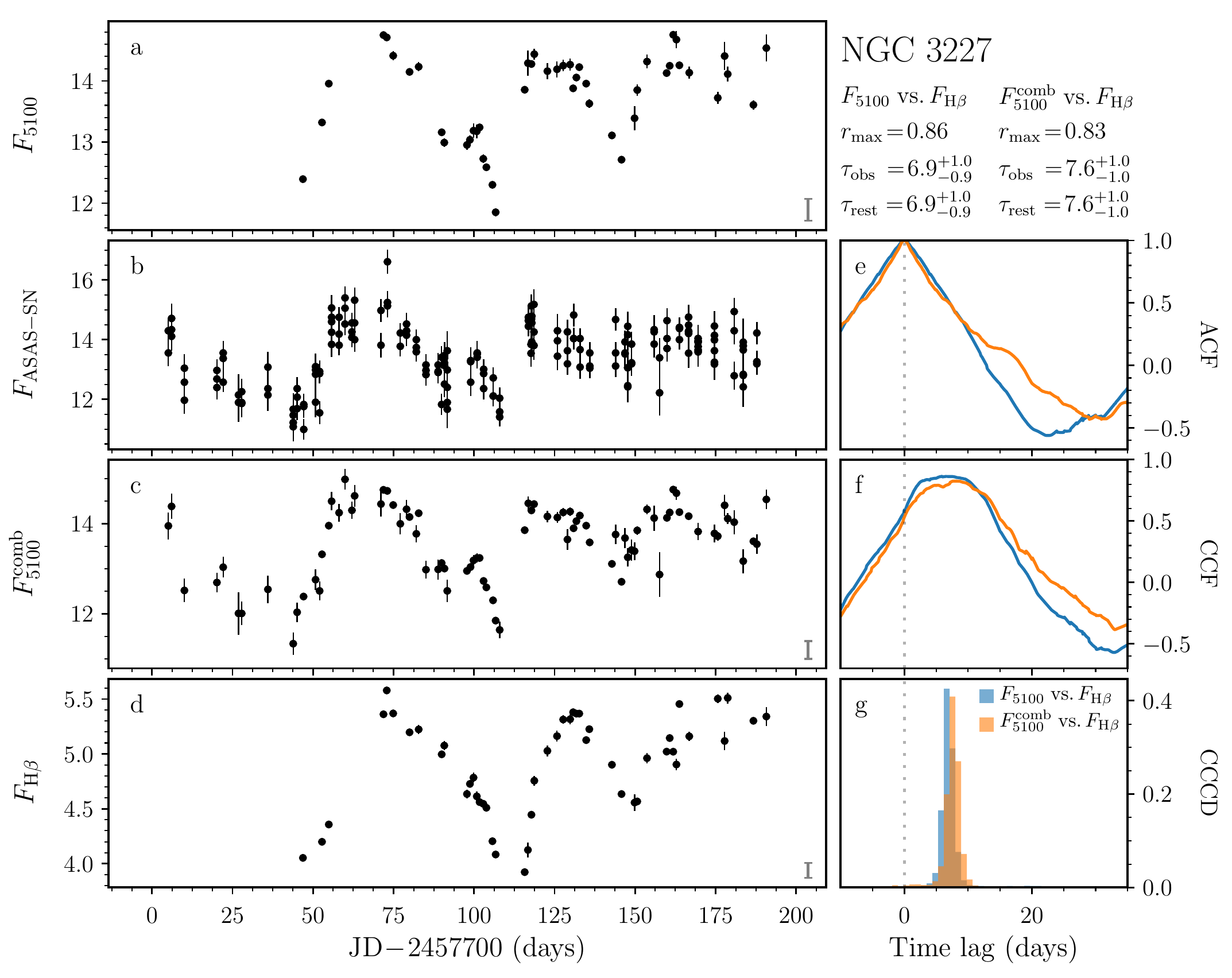} 
\caption{Light curves and cross-correlation functions. Panels a, b, c, and d are the continuum at 5100\AA, photometry from ASAS-SN,  combined continuum, and \hb\ light curves. Panels e, f, and g are the autocorrelation function (ACF), the cross-correlation function (CCF), and cross-correlation centroid distributions (CCCDs), respectively. We provide the name of the object and the corresponding lag measurements in both the observed and rest frames in the upper-right corner of the figures. The continuum and emission-line fluxes are ${10^{-15}\ \rm erg\ s^{-1}\ cm^{-2}\ \AA^{-1}}$ and ${10^{-13}\ \rm erg\ s^{-1}\ cm^{-2}}$, respectively. We show systematic errors, as the gray error bars in the
lower-right corners in panels a -- d (see details in Sections 2.2 and 2.3).}
\label{fig:light_curves} 
\end{figure*}

\begin{figure*}
\figurenum{\ref{fig:light_curves}}
\centering
\includegraphics[width=0.75\textwidth]{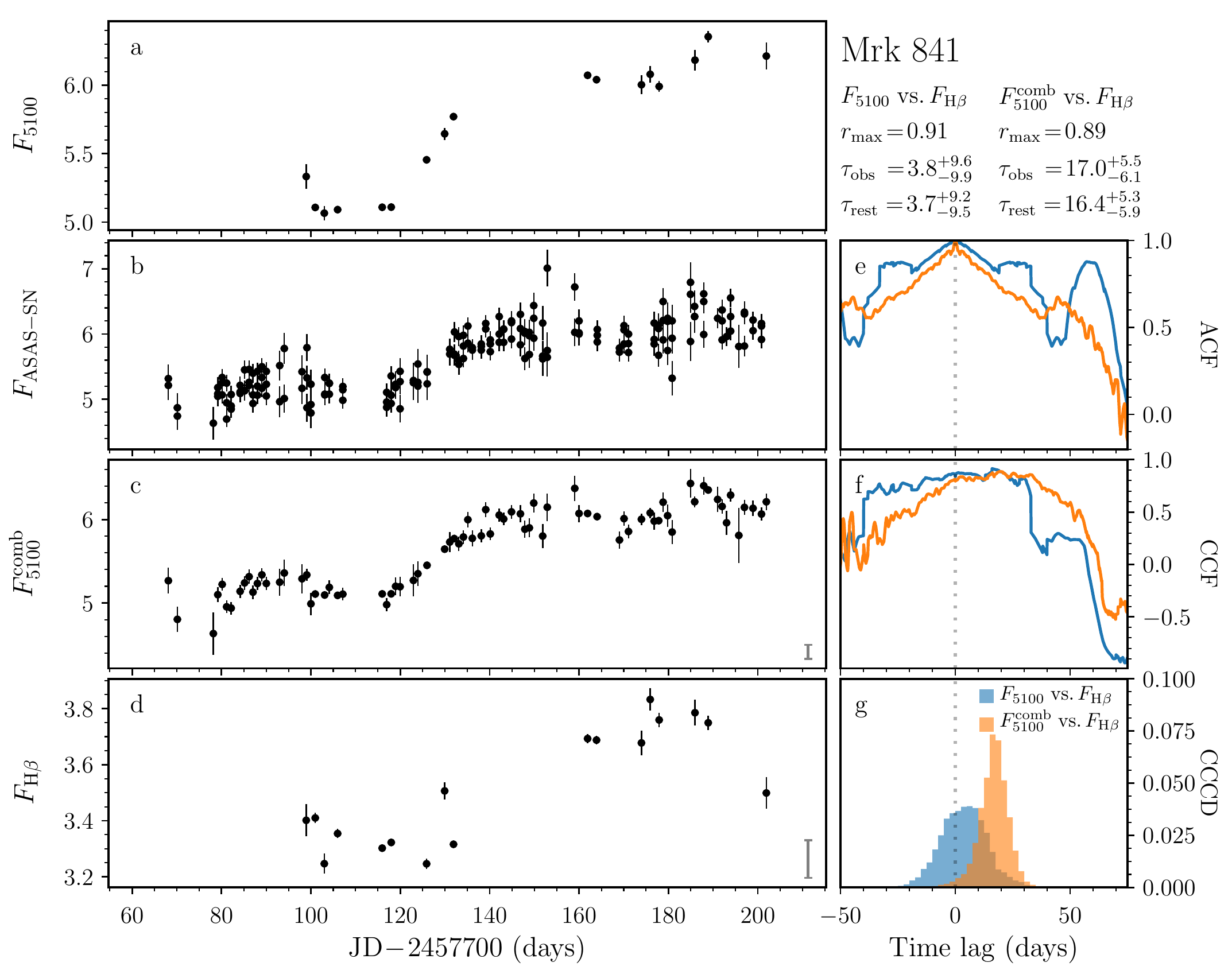} 
\caption{(Continued.)}
\end{figure*}

\section{Analysis}
\label{sec:analysis}

\subsection{Time Lag Determinations}
\label{sec:ccf}

For the first step in the  determination of the time delays between the continua 
and \hb\ emission lines, we used the interpolated cross-correlation function (ICCF; \citealt{gaskell1986, gaskell1987}).  Specifically, we take the centroid of the ICCF above 80\% of the peak (somewhat arbitrary, but a value used in many RM investigations, including our first MAHA paper) to measure the time line between two light curves.

We used the 
``flux randomization/random subset sampling (FR/RSS)'' method to estimate the uncertainties in the time delay
\citep{peterson1998, peterson2004}, a procedure that accounts for  both the measurement errors of the fluxes and the uncertainties due to the sampling/cadence. We simulate light curves consistent with our data set and its uncertainties, then use ICCF to produce cross-correlation centroid distributions (CCCDs) for the simulations. We adopted the median and 68\% confidence intervals of the CCCDs as our preferred time lags and their uncertainties.  Figure \ref{fig:light_curves} shows the auto-correlation functions (ACFs), 
the CCFs, and the CCCDs of the 5100\AA\ and \hb\ light curves for each object.
Table \ref{tab:lags} provides time-lag measurements and their corresponding 
maximum CCF correlation coefficients. We also provide  \hb\ time delays relative to the combined continuum light curves.

The scatter in the ASAS-SN light curves is generally larger than that in our WIRO continuum light curves, but it is worthwhile using the combined light curves when the advantages of the additional temporal sampling outweigh the disadvantages of the increased noise.  
 For NGC 3227, our WIRO 5100\AA\ continuum light curve has
homogeneous sampling and sufficient temporal extent. The time lags from WIRO or combined
WIRO-ASAS-SN continuum light curves are consistent within 1$\sigma$
uncertainties, and either may be used and yield quantitatively similar results.
For Mrk 79 and Mrk 841, however, the
advantages of the additional temporal coverage of ASAS-SN means tighter and more
symmetric CCCDs, with smaller uncertainties, so we adopted the results using the
combined continuum light curves. Table \ref{tab:lags} summarizes the CCF results
and indicates our adopted time lags and uncertainties, as indicated by a
``$\surd$''. For consistency we adopted the ASAS-SN light curves for all three objects.

%------------------------------ Table 4 ----------------------------------

\begin{deluxetable}{lcclclc}
%\rotate
  \tablecolumns{6}
  \setlength{\tabcolsep}{4pt}
  \tablewidth{0pc}
  \tablecaption{Time lags\label{tab:lags}}
  \tabletypesize{\scriptsize}%footnotesize}
  \tablehead{
      \colhead{}                         &
      \colhead{}                         &
      \colhead{}                         &
      \colhead{Observed}                 &
      \colhead{}                         &
      \colhead{Rest-frame}               &
      \colhead{}                         \\
      \colhead{Object}                   &
      \colhead{Continuum}                &
      \colhead{$r_{\rm max}$}            &
      \colhead{Time Lag}                 &
      \colhead{}                         &
      \colhead{Time Lag}                 &
      \colhead{Note}                     \\ 
      \colhead{}                         &
      \colhead{}                         &
      \colhead{}                         &
      \colhead{(days)}                   &
      \colhead{}                         &
      \colhead{(days)}                   &
      \colhead{}                         
            }
\startdata
Mrk~79       & 5100\AA  & 0.93 & $11.5_{- 3.4}^{+ 6.8}$ & & $11.3_{- 3.4}^{+ 6.6}$ &         \\
             & combined & 0.88 & $16.6_{- 3.6}^{+ 2.5}$ & & $16.2_{- 3.5}^{+ 2.4}$ & $\surd$ \\
NGC~3227     & 5100\AA  & 0.86 & $ 6.9_{- 0.9}^{+ 1.0}$ & & $ 6.9_{- 0.9}^{+ 1.0}$ &         \\
             & combined & 0.83 & $ 7.6_{- 1.0}^{+ 1.0}$ & & $ 7.6_{- 1.0}^{+ 1.0}$ &  $\surd$ \\
Mrk~841      & 5100\AA  & 0.91 & $ 3.8_{- 9.9}^{+ 9.6}$ & & $ 3.7_{- 9.5}^{+ 9.2}$ &         \\
             & combined & 0.89 & $17.0_{- 6.1}^{+ 5.5}$ & & $16.4_{- 5.9}^{+ 5.3}$ & $\surd$    
 \enddata
  \tablecomments{\footnotesize
  ``$\surd$'' means we use this time lag of the object to calculate its BH mass.}
\end{deluxetable}

\subsection{Line-width Measurements and Black Hole Masses}

Table \ref{tab:fwhm_mbh} provides our black hole mass determinations along with the associated H$\beta$ profile velocity measurements.  The black hole
masses are calculated using the virial equation:

\begin{equation}
\mbh = \fblr\frac{\rblr \Delta V^2}{G} = \fblr M_{\rm VP},
\end{equation}
where $\rblr=c\taublr$ is the emissivity-weighted radius of the BLR, $c$ is the speed
of light, $G$ is the gravitational constant, and  ($\Delta V$) measured from the FWHM ($V_{\rm FWHM}$) or line dispersion ($\sigma_{\rm line}$) \citep[e.g.,][]{peterson2004}. For $f_{\rm BLR}$, the empirically calibrated geometric correction factor multiplied by the virial product ($M_{\rm VP}$) to obtain mass, we adopt average values from \citet{woo2015}: 1.12 when using FWHM and 4.47 when using $\sigma_{\rm H\beta}$ from the rms spectra.  The value of $f_{\rm BLR}$ in individual objects likely differs 
due to effects like inclination and in many cases will represent the largest systematic source of uncertainty in derived masses \citep[e.g.,][]{pancoast2014b}.

%-------------------------------- Table 5 -------------------------------
\begin{deluxetable*}{lcccccccccc}%[h]
%\rotate
  \tablecolumns{6}
  \setlength{\tabcolsep}{3pt}
  \tablewidth{0pc}
  \tablecaption{\hb\ Width Measurements and Black Hole Masses\label{tab:fwhm_mbh}}
  \tabletypesize{\scriptsize}%footnotesize}
  \tablehead{
      \colhead{}                   &
      \multicolumn{2}{c}{mean spectra}   &
      \colhead{}                         &
      \multicolumn{2}{c}{rms spectra}      &  
      \colhead{}                         &
      \colhead{$M_{\rm VP}$ (rms spectra)}                    &
      \colhead{}                 &
      \multicolumn{2}{c}{$M_{\bullet}$ (rms spectra)}                 \\ \cline{2-3} \cline{5-6} \cline{8-8} \cline{10-11}
      \colhead{Object}                         &
      \colhead{$V_{\rm FWHM}$}                     &
      \colhead{$\sigma_{\rm line}$}           &
      \colhead{}                         &
      \colhead{$V_{\rm FWHM}$}                     &
      \colhead{$\sigma_{\rm line}$}           &
      \colhead{}                         &
      \colhead{$R_{\rm H\beta} V_{\rm FWHM}^2/G$}                    &
      \colhead{}                 &
      \colhead{$1.12\!\times\!R_{\rm H\beta} V_{\rm FWHM}^2/G$}                 &
      \colhead{$4.47\!\times\!R_{\rm H\beta} \sigma_{\rm line}^2/G$}                 \\
      \colhead{}                         &
      \colhead{(km s$^{-1}$)}            &
      \colhead{(km s$^{-1}$)}            &
      \colhead{}                         &
      \colhead{(km s$^{-1}$)}            &
      \colhead{(km s$^{-1}$)}            &
      \colhead{}                         &
      \colhead{($10^7 M_{\odot}$)}                         &
      \colhead{}                 &
      \colhead{($10^7 M_{\odot}$)}                         &
      \colhead{($10^7 M_{\odot}$)}                         }
\startdata
      Mrk~79  & $6191\pm78$ & $3505\pm32$ & &  $5087\pm27$ & $2092\pm18$ & & $8.19_{-1.77}^{+1.21}$ & & $ 9.17_{-1.98}^{+1.36}$ & $ 6.19_{-1.34}^{+0.92}$\\
%    NGC~3227  & $4032\pm57$ & $3370\pm33$ & &  $2476\pm21$ & $1502\pm24$ & & $ 0.83_{-0.11}^{+0.12}$ & & $ 0.93_{-0.12}^{+0.13}$ & $ 1.36_{-0.18}^{+0.20}$\\
    NGC~3227  & $4032\pm57$ & $3370\pm33$ & &  $2476\pm21$ & $1502\pm24$ & &  $ 0.91_{-0.12}^{+0.12}$ & & $ 1.02_{-0.14}^{+0.14}$ &  $ 1.50_{-0.20}^{+0.20}$\\
     Mrk~841  & $4993\pm45$ & $3393\pm16$ & &  $4011\pm26$ & $1852\pm24$ & & $ 5.15_{-1.86}^{+1.67}$ & & $ 5.77_{-2.08}^{+1.87}$ & $ 4.91_{-1.77}^{+1.59}$
 \enddata
  \tablecomments{We have corrected for the line spread function caused by the instrument and seeing.
  $M_{\rm VP}$ is the virial product measured from the rms spectrum. The black hole masses (\mbh) are calculated from the rms spectra using the \fblr\ factors from \citet{woo2015}. 
      }
\end{deluxetable*}

We measured the FWHM and the 
line dispersion $\sigma_{\hb}$ of the 
broad component of \hb\ for both the mean and rms spectra.  While velocities from the rms profile are most appropriate for mass determinations, values from the mean profile may also be useful and of interest. Our procedures largely eliminate the narrow-line region component of \hb\ from the rms spectra, but more care is required to remove narrow \hb\ from the mean spectra.  
We assumed that the profile of narrow \hb\ is the same as that of the [O III]
lines and subtracted off the narrow lines using the fitting procedure described by \citet{du2018b}.
Table \ref{tab:narrow_ratios} provides the narrow \oiii$\lambda5007$/\hb\ and \oiii$\lambda5007$/\oiii$\lambda4959$ flux ratios, which are close to typical values for AGNs (10 and 3, respectively, e.g., \citealt{kewley2006, stern2013}). Figure \ref{fig:meanrms} shows the narrow-line-subtracted \hb\ mean profiles.

%----------------------------- Table 6 --------------------------------

\begin{deluxetable}{lcc}
%\rotate
\tablecolumns{3}
\tablewidth{\textwidth}
\setlength{\tabcolsep}{5pt}
\tablecaption{Narrow-line flux ratios\label{tab:narrow_ratios}}
\tabletypesize{\footnotesize}
\tablehead{
\colhead{Object}                     &
\colhead{\oiii$\lambda$5007/\oiii$\lambda$4959}             &
\colhead{\oiii$\lambda$5007/\hb}                        
}
\startdata
Mrk~79  & 3.02$\pm$0.04 & 10.66$\pm$0.24 \\
NGC~3227& 3.03$\pm$0.03 & 10.21$\pm$0.43 \\
Mrk~841 & 3.04$\pm$0.04 & 12.85$\pm$0.64 
\enddata
%\tablecomments{      }
\end{deluxetable}

We determined the broad \hb\ velocity widths and their uncertainties using the bootstrap method. We measured FWHM and $\sigma_{\hb}$ for 500 resampled subset spectra for each mean/rms profile, where each resampled spectrum was formed by sampling $N$ data points from the $N$ data points in the corresponding original mean/rms spectrum. We used continuum and line windows appropriate for the entire mean profiles, not the truncated wavelength ranges based on the  rms profiles we used to create the \hb\ light curves. We adopted the median values and the standard deviations of the distributions of the 500 resampled spectra as our measurements and uncertainties, respectively. We empirically found a point spread function of 850-1000 km s$^{-1}$ (FWHM) for this WIRO slit width, grating, and campaign, as determined by \citet{du2018b}, and adopted an average width of 925 km s$^{-1}$, which we subtracted in quadrature from our line widths. This correction will contribute a modest additional systematic error, primarily in the AGNs with the narrowest broad H$\beta$ emission lines. 

\subsection{Velocity-resolved time lags}
\label{sec:velocity_resolved_lags}

\begin{figure}
\centering
\includegraphics[width=0.45\textwidth]{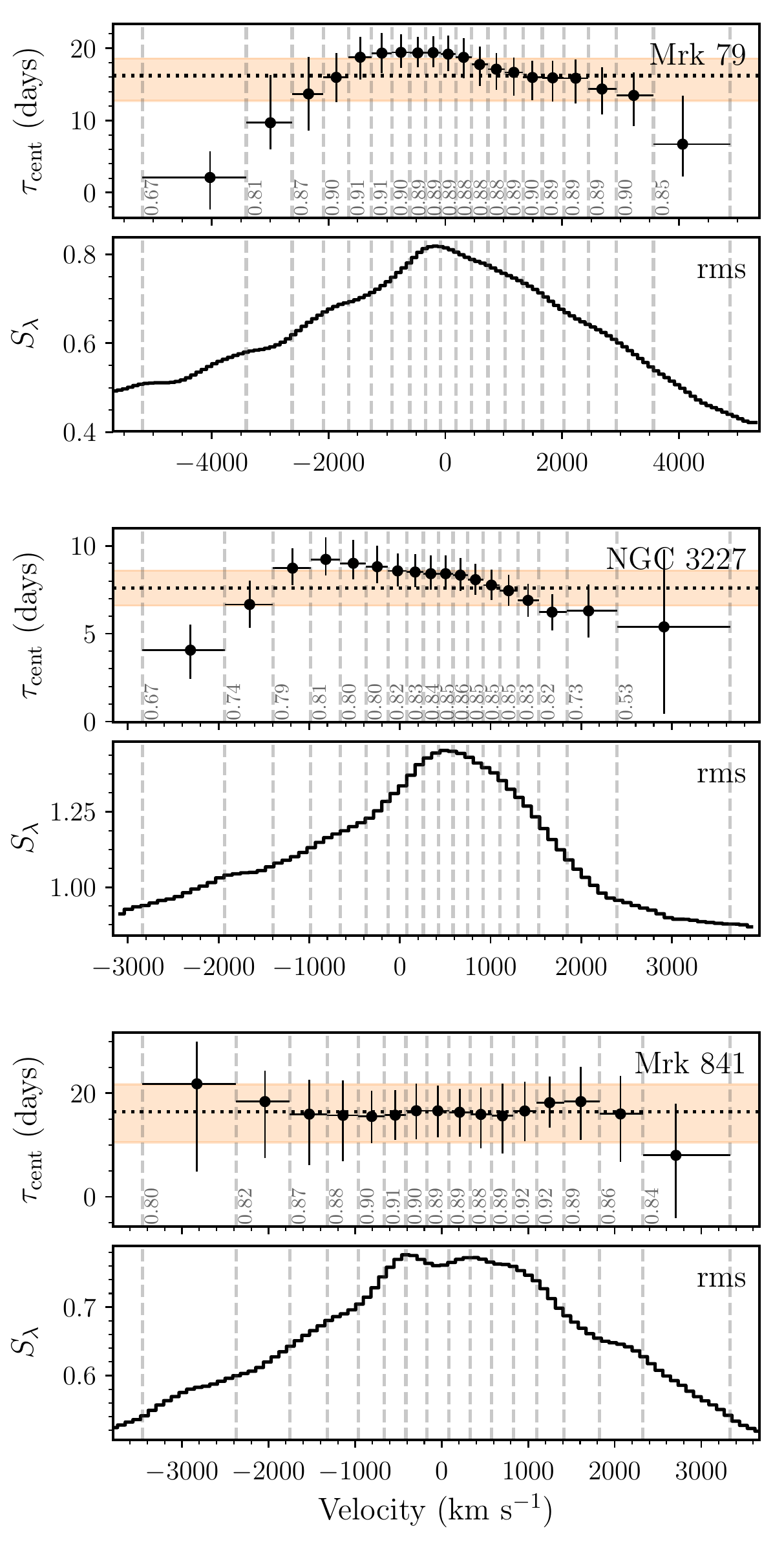}
\caption{Rest-frame velocity-resolved time lags and their corresponding rms spectra. The upper panels show
the centroid lags for each velocity bin as indicated by the vertical dashed lines. The vertical numbers inside each bin show the value of their peak correlation coefficient. The orange-shaded region shows the uncertainties (from Table \ref{tab:lags}) around the average time lags. The lower panels show the rms spectrum in units of ${10^{-15}\ \rm erg\ s^{-1} cm^{-2}\ \AA^{-1}}$.}
\label{fig:v_resolved_lags}
\end{figure}

To produce velocity-resolved time lags, we started with the continuum-subtracted rms spectrum, where Figure \ref{fig:meanrms} shows our adopted continuum windows. We separated each \hb\ profile into velocity bins of equal integrated flux. We then obtained
the \hb\ light curves for the bins and performed the same CCF analysis as previously (Section \ref{sec:ccf}) against each object's adopted 5100\AA\ continuum light curve. Figure \ref{fig:v_resolved_lags} shows the results.

\section{Two-Dimensional Velocity-Delay Maps (VDMs)}
\label{sec:2dvdms}

\begin{figure*}
    \centering
    \includegraphics[width=0.95\textwidth]{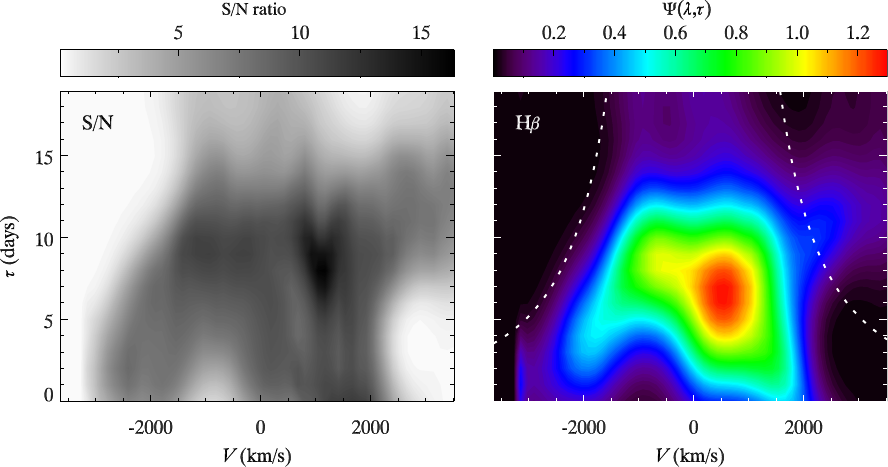}
    \caption{The velocity-delay map of NGC 3227 (right). The dotted lines indicate the ``virial envelope'' $V^2\tau c/G = 0.91\times10^7 M_{\odot}$, based on the ``virial product'' measured using the FWHM and the rms spectrum (see Table \ref{tab:fwhm_mbh}). The signal-to-noise (S/N) map (left). 
    }
    \label{vdm3227}
\end{figure*}

\begin{figure}
        \centering
        \label{fig:ngc3227-fit}
        \includegraphics[angle=0,width=0.45\textwidth]{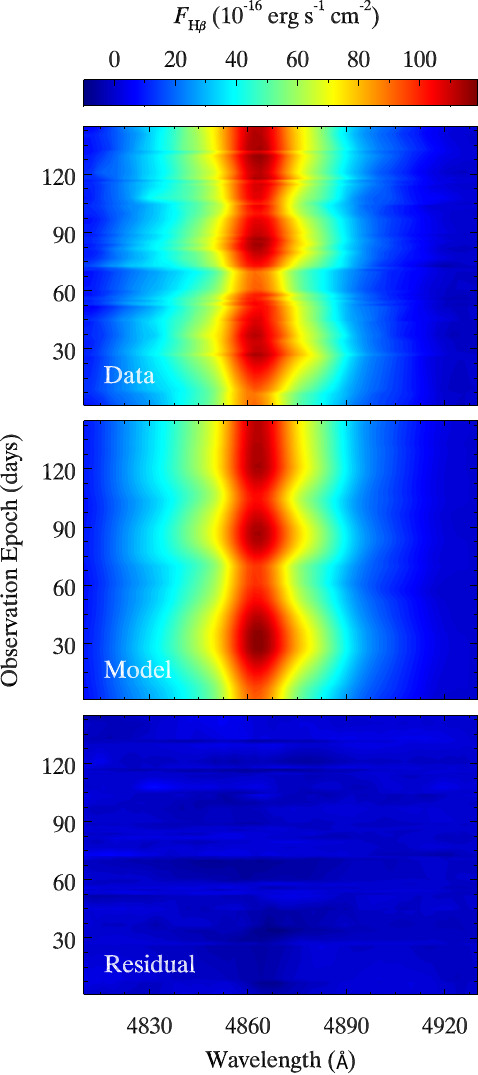} 
        \caption{\footnotesize 
                  The MEM fitting of the observed time series for the \hb\ profile of NGC 3227 and the residuals. 
        }
        \label{fit3227}
\end{figure}

The generally good data quality for Mrk 79 and NGC 3227 allows the calculation of two-dimensional VDMs using the maximum entropy method or MEM \citep[e.g.,][]{horne2004}.
Detailed descriptions of our approach were previously reported by \cite{xiao2018a,xiao2018b}, 
and we briefly introduce the principle here. 

MEM introduces a linearized echo model
$L(v_{i},t_{k})=\bar{L}_0(v_i)+\sum_{j}\varPsi(v_{i},\tau_{j})\left[C(t_{k}
-\tau_{j})-\bar{C}_0\right]\Delta \tau$
with a parameterized velocity-delay map $\varPsi(v_{i},\tau_{j})$, to fit the varying emission-line profile $L(v_{i},t_{k})$, continuum light curve $C(t_{k})$, and the non-varying emission-line background $\bar{L}_0(v_{i})$ simultaneously. We adopt the median of the continuum data as the reference continuum level $\bar{C}_0$. The MEM fitting uses ${\chi}^{2}=\sum_{m}\left[D_{m}-\mathscr{M}_{m}(\vec{p})\right]^{2} /\sigma_{m}^{2}$ to measure the differences between the model $\mathscr{M}_{m}(\vec{p})$ and the data $D_{m}$, and to control the ``goodness-of-fit.'' Complementary to this, the ``simplicity-of-modeling" is achieved by introducing the entropy $S=\sum_{n}\left[p_{n}-q_{n}-p_{n}\ln(p_{n}/q_{n})\right]$ to constrain the model parameters $p_{n}$ by the ``default image'' $q_{n}$ (e.g. $q(t)=\sqrt{p(t-\Delta t)p(t+\Delta t)}$ for the one-dimensional $C(t)$). The MEM fitting uses a user-controlled parameter $\alpha$ to keep balance between the ``goodness-of-fitting'' and the ``simplicity-of-modeling", and is accomplished by minimizing $Q={\chi}^{2}-\alpha S$. 

In order to make an estimation of the uncertainties, we 
followed previous efforts \citep{grier2013, xiao2018a} and 
first used a damped random walk (DRW, \citealt{li2013, zu2013}) to model the
continuum (shown in Appendix \ref{app:recon_LC}), and then adopted
the resulting highly sampled continuum to replace the original in the
fitting.  \cite{xiao2018a} demonstrated that using a  linear
interpolation or the original continuum light curves in the MEM does not change
the output velocity-delay map significantly.

We produced signal-to-noise ratio maps by using a flux randomization scheme
\citep{peterson1998}.
We created 100 alternative but statistically consistent values for each datum and used these to create new VDMS.  From this set of 100 maps, we computed the corresponding signal-to-noise ratio map.  See \citet{xiao2018b} for additional details.

We plot the VDM and associated MEM fitting of the observed time series for the \hb\ profile of NGC 3227 in Figures 
\ref{vdm3227} and \ref{fit3227}, respectively. We do the same for the more complex case of Mrk~79 in Figures \ref{vdmmrk79} and \ref{fit79} and provide some additional work described below in the next section to help interpret the resulting VDM.

\subsection{An illustrative model for Mrk 79}
\begin{figure*}
    \centering
    \includegraphics[width=0.95\textwidth]{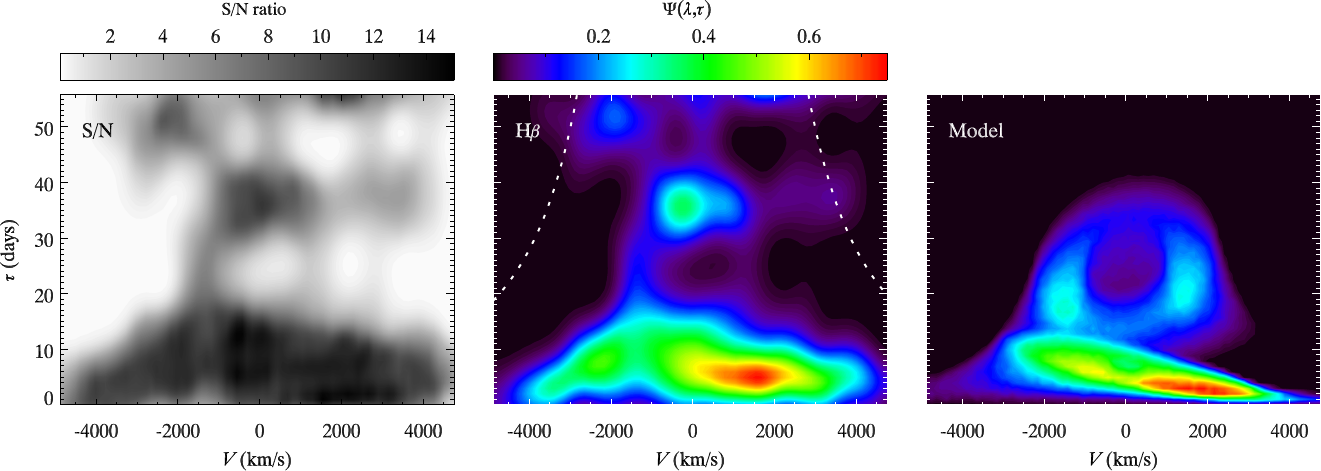}
    \caption{The velocity-delay map of Mrk 79 (middle). The dotted lines indicate the ``virial envelope'' $V^2\tau c/G = 8.19\times10^7 M_{\odot}$, based on the ``virial product'' measured using FWHM and the rms spectrum (see Table \ref{tab:fwhm_mbh}). The signal-to-noise (S/N) map (left). The velocity-delay map of the illustrative model for Mrk 79 (right). The BLR here is composed of an outflowing disk at larger distances and an inflowing disk at smaller distances. The parameters of the simulation are given in Table \ref{tab:simuparameter}.
    }
    \label{vdmmrk79}
\end{figure*}

\begin{figure}
        \centering
        \label{fig:Mrk79-fit}
        \includegraphics[angle=0,width=0.45\textwidth]{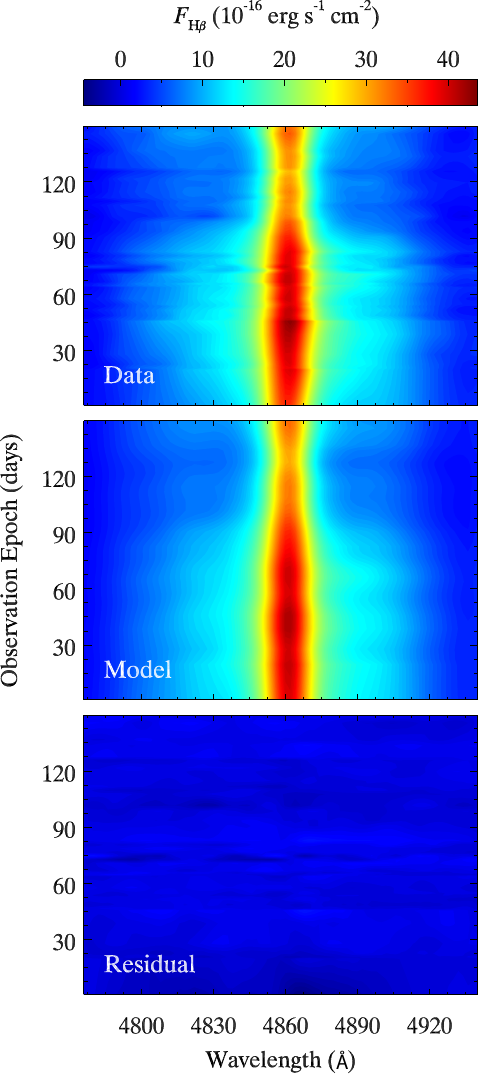} 
        \caption{\footnotesize 
                   The MEM fitting of the observed time series for the \hb\ profile of Mrk 79 and the residuals. 
        }
        \label{fit79}
\end{figure}

Interpreting VDMs is not always
straightforward, especially in cases like that of Mrk 79.  The VDM of Mrk 79 shows something that looks like it may possess multiple dynamical components and hot spots.  We did not perform an extensive search of parameter space, something which is computationally challenging and without an established approach, but did make an attempt to create a simple multi-component simulation to reproduce a VDM with similar features as those seen in Figure \ref{vdmmrk79}. The upright ellipse is suggestive of a disk component, while the short-time-lag emission may arise from another disk closer to the continuum source. These ideas form the starting point for one possible way to quantitatively interpret the Mrk 79 VDM.  Interactive adjustment of model components can lead to a potentially interesting model that may provide insights about the nature of the reverberating BLR.

Figure \ref{vdmmrk79} (right) provides 
a VDM of a model BLR that has some resemblance to that of Mrk~79.
This simulation shows the combined signal from two separated gas disks in the BLR. The kinematics of one disk is dominated by inflow, the other outflow. The inflowing disk is close to the central BH, while the outflowing one is located at larger distances. We adopt the FWHM-based black hole mass measured from this campaign (See Table \ref{tab:fwhm_mbh}) in the simulation, and used different parameters of the inner and outer radii ($R_{\rm in}$ and $R_{\rm out}$), the radial and tangential velocities of the BLR clouds ($V_r$ and $V_{\Phi}$) as fractions of the virial velocities, the inclination angle and the opening angle ($\theta_{\rm in}$ and $\theta_{\rm o}$), and the spatial emissivity distribution factor $\gamma$ of the BLR clouds ($R^{\gamma}$) for the two disks. The values of these parameters are listed in Table \ref{tab:simuparameter}.
The magnitude of the velocity of one BLR cloud at a distance $r$ is $(V_r^2+V_{\Phi}^2)$, where again the values of $V_r$\ and $V_{\Phi}$\ describe the fractions of the local virial velocity, and positive/negative $V_r$\  indicate outflow/inflow.

        \begin{deluxetable*}{cccccccccc}
                \tablecolumns{11}
                \tablewidth{0pc}
                \tablecaption{Values of the parameters for the double-disk model shown in Figure \ref{vdmmrk79}. %\ref{fig:simu}.
         \label{tab:simuparameter}}
                \tabletypesize{\footnotesize}
                \tablehead{
                        \colhead{}                  &
                        \colhead{}                  &
                        \colhead{$M_{\rm BH}$}      &
                        \colhead{$R_{\rm in}$}      &
                        \colhead{$R_{\rm out}$}     &
                        \colhead{$V_{\rm r}$}  &
                        \colhead{$V_{\rm \Phi}$}      &
                        \colhead{$\theta_{\rm in}$} &
                        \colhead{$\theta_{\rm o}$}  &
                        \colhead{$\gamma$}          \\
                        \colhead{}                  &
                        \colhead{}                  &
                        \colhead{$10^{7}(M_{\odot})$}                &
                        \colhead{(ltd)}                &
                        \colhead{(ltd)}                &
                        \colhead{$V_{\rm virial}$}                &
                        \colhead{$V_{\rm virial}$}                  &
                        \colhead{(deg)}                &
                        \colhead{(deg)}                &
                        \colhead{}                  
                }
                \startdata
                 Inflowing disk          &  &9.17        &0.1    &8      &$-$0.4      &0.3    &60     &25          &1   \\
                Outflowing disk         &  & ...         &12     &28     &0.2        & 1  &67     &10          &0.6
                \enddata
        \end{deluxetable*}

While the general shape is similar between the simulated VDM and that of Mrk~79 in our campaign, the emissivity pattern differs.  Matching them more closely would require inhomogeneous gas distributions in the disks.  
For example, the response at $-$4000 to $-$2000 km/s (delay $\sim$5 days) can be explained 
if the radiation of the outflowing disk has inhomogeneity and the near side is brighter than 
its far side.
We will further discuss the interpretation of the Mrk~79 VDM below.

\section{Discussion}
\label{sec:discussion}

We have measured both integrated \hb\ time lags and velocity-binned
\hb\ time lags for Mrk~79, NGC~3227, and Mrk~841.  The integrated lags may be simply interpreted as the emissivity-averaged radius of the varying \hb\ BLR during the campaign, which we have used to estimate the masses of each target's central black hole mass assuming virial motions.  This is a standard assumption and leads to a conventional analysis \citep[e.g.,][]{peterson2004}.

The velocity-resolved time lags provide
more information and can potentially unlock the kinematics of the BLR, but can be challenging to interpret.  Emission-line gas contributing to each bin is not likely to be co-spatial, as Doppler shifts are for the velocity component along the line of sight only.  
Two-dimensional velocity delay maps 
(VDMs) have been computed for particular kinematic models in past investigations for the purpose of comparison to maps generated to match observational data
\citep[e.g.,][]{welsh1991,horne2004,bentz2009, wang2018}.  If a successful VDM can be found, in principle it can break degeneracies in the time lag for a given velocity and reveal two-dimensional structures diagnostic of the BLR kinematics.  The VDMs can also sometimes be challenging to interpret. We discuss our results for each object individually below, including a comparison to past RM investigations as appropriate.  We note that detailed corrections for residual host-galaxy contamination will be left for a future project.

\subsection{Individual Objects}\label{sec:individual}

\subsubsection{Mrk~79}
\label{sec:mrk79}
We first compare our results to two previous reverberation mapping campaigns \citep{peterson1998,lu2019}.  Due to some differences between object and background apertures as well position angles, it is not totally straightforward to compare in detail the flux levels of our campaign with these previous investigations.  The 5100\AA\ continuum during our campaign ranged between about 2.5-3.5
$\times$10$^{-15}$ ergs s$ ^{-1}$ cm$^{-2}\ {\rm \AA}^{-1}$,
not very dissimilar to the range of 
2.7-3.5 $\times$10$^{-15}$ ergs s$ ^{-1}$ cm$^{-2}\ {\rm \AA}^{-1}$ for that of \citet{lu2019}, who used a narrower slit of 2.5 arcseconds, and whose observations were the season immediately after ours.  These recent flux levels are approximately half as bright as during the 1990s when \citet{peterson1998} first detected a time lag for Mrk~79.

The 1990s season with the best data 
and strongest variations (Julian Days 2448193-2448393) showed an \hb\ lag of  $\tau_{\rm cent}=18.1_{-8.6}^{+4.9}$ days. Updated analyses give a slightly shorter measurement of about 16 days rest-frame with an uncertainty of 35-40\% \citep{bentz2013}.  Using our adopted f factor for $\sigma_{line}$, their corresponding black hole mass estimate for Mrk 79 is 4.3 $\pm 1.2 \times 10^7 M_{\odot}$.

Much more recently, \citet{lu2019} reported their results for an RM campaign in the year just afterour observations.  They found a Mrk 79 \hb\ time lag of 3.49$^{+0.62}_{-0.60}$ days in the rest-frame, very much shorter than that of \citet{peterson1998} and with a much smaller uncertainty.  The 1990s campaign had a relatively poor cadence and would have been challenged to accurately measure such a short time lag. The \citet{lu2019} velocity-resolved time lags showed the shortest delays (about 2 days) at the most blueshifted velocities, and longer 
time lags at redshifted velocities, suggestive of an outflow.  They determined a black hole mass of 4-5$\times$10$^7$ M$_\odot$
depending on which $\Delta V$ measurement they used, with an uncertainty of approximately 30-40\%.  While that uncertainty is relatively large, it is for a very small time lag that clearly differs from the 1990s lag measured by \citet{peterson1998},
but the mass estimates are similar.

Our RM campaign showed a rest-frame centroid time lag of $16.2_{- 3.5}^{+ 2.4}$ days, very similar to that of \citet{peterson1998}, but a somewhat a larger mass (6-9 $\times$10$^7$ M$_\odot$ depending on which $\Delta V$ was used).  We will focus on comparing
our results with those of \citet{lu2019}, which produced velocity-resolved time lags and was close in time to our campaign.  Consulting our VDM, we see significant emissivity at a few days time lag, also consistent with that of \citet{lu2019}, but with a pattern more consistent with inflow rather than outflow.  The reason we have such a longer time lag, closer to that of \citet{peterson1998}, is the additional response we see at much longer timescales.

The largest change in \hb\ in our
light curve is the drop starting at around Julian Day2457840, larger in amplitude than the previous variations, and we have an additional two months of data after that.  Moreover, the continuum has a general downward trend until the very end of our campaign.  Any small changes can be seen in the most quickly responding gas, but larger steadier changes are required to see a large variation on larger size scales. 
Short and small variations will be washed out on large scales, and the \citet{lu2019} light curve does not have such a large consistent drop, but rather a series of up and downs on $\sim$20-day time scales.  We suggest that the \citet{lu2019} campaign saw continuum variations best suited to highlight gas at small radii, while ours was suited to see gas at both small and large radii.  \citet{goad2014} have previously explored such issues as this geometric dilution effect for an extended BLR.  Responses at larger radii are effectively smoothed out.

To explain the general appearance of the VDM of Mrk 79, our illustrative model has two gas disks with 
different inner and outer edges (in other words, two thick rings) at two distances in the BLR with similar and relatively high inclination angles for type 1 AGNs
(60 and 67 degrees).  It is important to keep in mind that RM is only picking up \hb\ gas that changes in response to the continuum light curve.  These two components may simply reflect the migration of a single \hb\ emitting disk to smaller radii as the continuum drops, and the largest changes are at the inner and outer edges.  Gas at intermediate distances may simply not change its emissivity very strongly, for example if the ionization state does not change significantly or if the intermediate distance material is optically thin 
\citep{shields1995}. Our simulated disks suggest another possibility, with the inner disk having a much larger opening angle of 25 degrees compared to the 10 degrees of the outer disk.  Perhaps a large covering fraction of optically thick clouds in the inner disk filters the ionizing continuum to the region intermediate between our two components.
It would be of interest to generate
VDMs of lines of different ionization to see if the full disk can be mapped out to test these ideas.

Mrk 79 presents a case in which different high-quality RM campaigns  have found very different \hb\ time lags at different epochs but not very dissimilar luminosities.  We have no reason to believe that both these results are not real, and that different campaigns with different light curves, cadences, and durations can be sensitive to responses in different parts of the BLR.  This is likely a general feature of BLR RM and a source of intrinsic scatter in scaling relationships.  Different parts of a disk may dominate the reverberations during different times.  

\citet{afanasiev2019} report a spectropolarimetry-based mass for Mrk 79 of log ($M_{\bullet}$/$M_{\odot}$) = 7.45$\pm$0.27, or 2.8$\times$10$^7$ M$_\odot$, about a factor of two lower than the RM-based masses, albeit with a moderate uncertainty (a factor of 1.9).  
We note that they found a more edge-on inclination angle of 49.1 $\pm$ 9.8 degrees, similar to the somewhat larger angles we used for our model disks.  We continue to see differences of factors of at least a few in the mass determinations using RM measurements and other new techniques.  Higher quality data allowing more detailed analysis and deeper understanding of the BLR and its changes may help resolve these issues.  

\subsubsection{NGC~3227}
\label{sec:ngc3227}

NGC~3227 has been spectroscopically monitored many times in past decades by different groups
\citep{salamanca1994, winge1995, onken2003, denney2009, denney2010, derosa2018}.
We will first discuss the last two high-quality
RM campaigns, which produced velocity-resolved time lags, and then compare them to our own results.  

\cite{denney2010} reported that in 2007, when the 5100\AA\ continuum flux was  $\sim3.2-5.1\times$10$^{-15}$ ergs s$ ^{-1}$ cm$^{-2}\ {\rm \AA}^{-1}$ (with a weighted mean of $3.27\times$10$^{-15}$ ergs s$ ^{-1}$ cm$^{-2}\ {\rm \AA}^{-1}$) after subtracting the host contribution, the \hb\ centroid lag was
$3.75_{-0.82}^{+0.76}$ days in the rest frame, and for the first time they measured 
velocity-resolved time lags for NGC 3227.  The red wing showed longer time lags than the blueshifted emission, suggestive of an outflow.  Moreover, during that epoch the mean \hb\ line showed a mild red asymmetry, along with a complex and double-peaked rms profile.

More recently, \citet{derosa2018} report results from an RM campaign in 2012 and 2014 when the continuum flux ranged over$\sim$13-16$\times$10$^{-15}$ ergs s$ ^{-1}$ cm$^{-2}\ {\rm \AA}^{-1}$ (with a mean of 15.6$\times$10$^{-15}$ ergs s$ ^{-1}$ cm$^{-2}\ {\rm \AA}^{-1}$). If subtracting the host contribution of $8.6 \times 10^{-15} \ \mathrm{erg\ s^{-1}\
cm^{-2}\ \AA^{-1}}$ in their $5^{\prime\prime} \times 12^{\prime\prime}$
aperture \citep{derosa2018}, the AGN flux is $7.0 \times 10^{-15} \ \mathrm{erg\ s^{-1}\ cm^{-2}\
\AA^{-1}}$.
The average time lags were also very short, from $\sim$1-3 days in the rest-frame during these epochs, 
with a JAVELIN measurement of 2.29$^{+0.23}_{-0.20}$ days for
the 2012 campaign, shorter
than during 2007 despite a somewhat larger luminosity.
The mean \hb\ profile
then had changed to a single-peaked line
with a blue asymmetry (at least in 2012).
While the time lags at the most extreme velocities were more uncertain, it appeared that the signatures of outflow had changed to those of inflow, with a blueshifted time lag peak and shorter time lags at redshifted velocities.

In 2017 we monitored NGC 3227 when it
still had a similar mean continuum flux $13.8 \times
10^{-15} \ \mathrm{erg\ s^{-1}\ cm^{-2}\ \AA^{-1}}$, but found a longer
time lag of $7.6_{-1.0}^{+1.0}$ days in the rest frame. 
Our campaign used an aperture
($5^{\prime\prime} \times 12.6^{\prime\prime}$) almost as same as \cite{derosa2018}, 
thus after subtracting the host contribution, the AGN flux in our
campaign is $5.2 \times 10^{-15} \ \mathrm{erg\ s^{-1}\ cm^{-2}\ \AA^{-1}}$. 
Given the radius--luminosity relation ($R_{\rm H\beta}-L_{5100}$ relation) of a single object (with a slope of
0.79) deduced from NGC 5548 in \cite{kilerci_eser2015}, the $R_{\rm
H\beta}-L_{5100}$ prediction in our campaign based on the time lag
and flux in \cite{denney2010} should be $\sim5.4$ days. Our time lag ($7.6_{-1.0}^{+1.0}$ days) is
marginally consistent with this prediction within the 2$\sigma$ uncertainties. The lag measurement in \cite{derosa2018} 
is significantly shorter than the prediction based on the $R_{\rm
H\beta}-L_{5100}$ (even using the traditional relation in \citealt{bentz2013}) and the measurements in
\cite{denney2010}.

During our observations,
the \hb\ line profile showed only a minor blue asymmetry, as did the rms spectrum, although the peak of the latter was redshifted by 500 km~s$^{-1}$).
Our velocity-resolved time lags (Fig. 3) qualitatively agree with those of \citet{derosa2018}, with the longest lags being at blueshifted velocities.  The highest-velocity bins, both blueshifted and redshifted, show short time lags of around 5 days compared to the longer time lags of the low velocity bins.
Our VDM (Fig. \ref{fig:ngc3227-fit}) shows \hb\ emission 
filling the expected virial envelope, with less on the blue side than the red.  This may represent an extended disk with an inflowing component and some obscuration, although there is not a lot of detail in the map.

 The changes of the average time delay are not easy to interpret in our conventional understanding as they are not tied to the luminosity changes.
There is a change in the \hb\ profile that corresponds to a change in the kinematics.  The double-peaked red asymmetric \hb\ profile of 2007 displayed an apparent outflow, while the mildly blue asymmetric \hb\ seen in later epochs showed elements of inflow.  A disk-like component may be present at all times, but the asymmetry may be connected to the presence of a component of radial flow.  Extrapolating the 
results for NGC 3227 probably a bit too far,
red asymmetric profiles may indicate outflow, while blue asymmetric profiles may indicate inflow here.  This could make sense if the \hb-emitting gas were at least partially opaque, making the far side of the BLR brighter than the near side.  Before coming to firmer conclusions a much larger data set of high-fidelity RM results should be compiled and analyzed statistically, and there exist counter examples (e.g., see Mrk 817 and NGC 3516 as reported by \citealt{denney2010} and \citealt{derosa2018}).  Perhaps there is no simple
connection between line asymmetry and
kinematics, but the statistics are so 
far rather small. 

Even if the average time lags and the kinematics of the \hb\ line have changed
over the decade considered, the campaigns should yield consistent mass measurements if the gas is moving virially.  Deviations from gravitational kinematics could be associated with the presence of radial motion, for instance, in the case of radiation-driven outflows
\citep[e.g.,][]{krolik2001}.  A consistent comparison
can be done using the line dispersion $\sigma_{\rm line}$ for the \hb\ profile
velocity width.  All campaigns measured 1400-1500 km s$^{-1}$.  The first two RM
programs get similar masses of about 0.6 $\times 10^7 M_{\odot}$ (for
consistency using an updated f factor for \citealt{denney2010}), but our present
effort is about a factor of two higher (1.50$\times 10^7M_{\odot}$)  due to the longer time lag. 
This is suggestive that there may be
non-virial motions present in the \hb-emitting BLR gas at some epochs, but other
explanations may also be possible.  For instance, the surrounding BLR may evolve
over time, with changing distributions of gas that could lead to a different
average distance of emitting clouds even with the same luminosity.

NGC 3227 also has several black hole mass measurements using techniques other than RM.  Using the spectropolarimetry method,
\citet{afanasiev2019} found a black hole mass of log ($M_{\bullet}$/$M_{\odot}$) = 7.31$\pm$0.22, or 2.04$\times$10$^7$ $M_{\odot}$ with an uncertainty $\sim$66\%.  This is larger than our mass, but marginally consistent within the uncertainties.
Almost uniquely among Seyfert 1 galaxies with good RM campaigns, NGC 3227 has black hole masses based on both gas and stellar dynamics.
The gas dynamics observation 
provides a measurement of $2.0^{+1.0}_{-0.4} \times 10^7$ $M_{\odot}$ \citep{hicks2008}, consistent with the higher mass.  The stellar dynamics measurement is 0.7 to 2.0
$\times 10^7 M_{\odot}$ \citep{davies2006}, 
inclusive of both the higher and lower
masses found.
The various mass determinations range from 0.6 to 2 $\times$10$^7$ $M_{\odot}$ and it is likely we know the mass of NGC 3227's central black hole to within a factor of three or so, all these different techniques taken together.

\subsubsection{Mrk~841}
\label{sec:mrk841}

While our dataset for Mrk 841 is the poorest
of the three objects investigated in this paper,
we did successfully measure a time lag.
The time lag has substantial uncertainty 
due to a gap during March 2017
caused primarily by bad weather.  At some point in that gap the H$\beta$ line stepped up in flux but we missed the precise dates of that transition.  

This object does not have a time lag 
reported in the literature, although there 
has been at least one previous attempt.
\citet{barth2015} report observing Mrk~841 in an RM campaign in 2011, but due to weak variability failed to measure a clear time lag.  
We enjoyed large variability and managed to obtain a time lag measurement, aided significantly by the ASAS-SN continuum light curve.

The \hb\ profile in the rms spectrum of Mrk~841 showed was roughly symmetric during our campaign. Its average \hb\ time lag was 
$16.4_{-5.9}^{+5.3}$ days in the rest frame. 
However, the \hb\ lags at different velocities were not readily distinguishable (see Figure \ref{fig:v_resolved_lags}) despite good velocity resolution.  
Our uncertainties were relatively large, $\sim$35\%, and perhaps this hid real differences.  
Another possible interpretation  of the 
relatively flat time lag response across the profile is that the \hb-emitting BLR region at this epoch was a not very extended ring seen at a small viewing angle. If we take the slight rise in time lag in the bluest bin at face value, and the drop in the extreme red bin, then some degree of inflow is suggested.

Our RM-derived virial mass is $\sim$3-7$\times10^7 M_{\odot}$ or 
log $M_{\bullet}$ = 7.5-7.8, considering
our uncertainties and the two velocity width
measurements and calibrations.  Recently
\citet{afanasiev2019} reported a black
hole mass of log ($M_{\bullet}$/$M_{\odot}$) = 8.76$\pm$0.27, an order of magnitude larger than
our result.  Their spectropolarimetry method is in principle largely independent of 
inclination angle and could prove to be 
a good alternative to more resource intensive RM methods, and 
generally agrees with RM-based black hole mass estimates much better than in  this instance \citep{afanasiev2015}.

\section{Future Work}
This paper has focused only upon the \hb\ line and its response and the inferred BLR kinematics for three objects observed during our first MAHA campaign.  Future investigations for these objects, as well as other MAHA targets, will include:

\begin{itemize}
    \item{Results of continued monitoring of these three objects in follow-up campaigns.}
    \item{A more objective and comprehensive technique for searching parameter space to produce models of VDMs.}
    \item{Dynamical modeling of Mrk 79 and NGC 3227. Dynamical modeling seeks to reproduce a fully three-dimensional representation of the BLR, and some reasonably successful results have been obtained for some objects \citep[e.g.,][]{pancoast2012,pancoast2014b,grier2017,li2018,pancoast2018,williams2018}.  This technique can provide information about system orientation and the black hole mass independent of other calibrations that possess their own uncertainties.}
    \item{The examination of other emission lines, particularly He II $\lambda$4686, which generally has shorter time lags than \hb\ and can provide information about the innermost BLR.}
    \item{Placement of MAHA objects on the radius-luminosity relationship, and looking for deviations associated with \hb\ asymmetry and/or kinematics.}
\end{itemize}

\section{Conclusions}
\label{sec:summary}

A few general conclusions emerge from this and some other recent high-fidelity RM campaigns.  Integrated emission-line time lags only tell part of the story,
and can change dramatically over time.  
Emission-line profiles and 
BLR kinematics, as characterized through velocity-resolved time lags and VDMs, can
change over the  course of a few years or less in Seyfert galaxies.  There may not be a a single BLR kinematic structure,
at least not uniformly seen in reverberating gas, and individual objects may have rather complex BLRs (e.g., Mrk 79).

During December 2016 through May 2017:

\begin{enumerate}

\item{Mrk 841: The rest-frame time lag of the \hb\  line was $16.4_{-5.9}^{+5.3}$ days. 
The black hole mass is likely between 10$^7 M_{\odot}$ and 10$^8 M_{\odot}$. 
We did not measure strong variations of time lag with velocity, perhaps because 
of the relatively large uncertainty on our time lags for this target due to gaps in the spectroscopic monitoring.  Slightly shorter time lags at the most redshifted velocites and slightly longer at the most blueshifted velocities might indicate inflow, but more investigation is required.}

\item{NGC 3227: The rest-frame time lag of the \hb\  line was $7.6_{-1.0}^{+1.0}$ days. 
The black hole mass is on the order of 10$^7 M_{\odot}$  The velocity-resolved time lags are most easily interpreted as arising from a BLR featuring a disk with an inflowing component.}

\item{Mrk 79:  The rest-frame time lag of the \hb\  line was $16.2_{- 3.5}^{+ 2.4}$ days. 
The black hole mass is most likely 6-9 $\times$ 10$^7 M_{\odot}$.  The velocity-resolved time lags are complex, with perhaps both inflowing and outflowing components on different size scales that are the changing parts of a single large disk.}

\end{enumerate}

\acknowledgments

We acknowledge the support by the National Key R\&D
Program  of  China  (grants  2016YFA0400701  and
2016YFA0400702),  by  NSFC  through  grants  NSFC-11873048,  -11973029,  -11721303, -11833008, -11991052, -11991054, -11503026, -11233003, -11573026, -11773029, by the Strategic Priority Research Program of the Chinese
Academy of Sciences Grant No.  XDB23000000, and by
Grant No.  QYZDJ-SSW-SLH007 from the Key Research Program of Frontier Sciences, CAS. 
Michael Brotherton enjoyed support from the Chinese Academy of Sciences President’s International Fellowship Initiative, Grant No. 2018VMA0005.
We also acknowledge support from a University of Wyoming Science 
Initiative Faculty Innovation Seed Grant.  
We thank Henry Kobulnicky and WIRO engineer James Weger for their invaluable assistance.
This research has made use of the NASA/IPAC Extragalactic Database (NED), which is operated by the Jet Propulsion Laboratory, California Institute of Technology, under contract with the National Aeronautics and Space Administration.

\appendix

\section{Some examples of the flux-calibrated spectra}
\label{app:spectra}

\begin{figure*}
    \centering
    \includegraphics[width=0.95\textwidth]{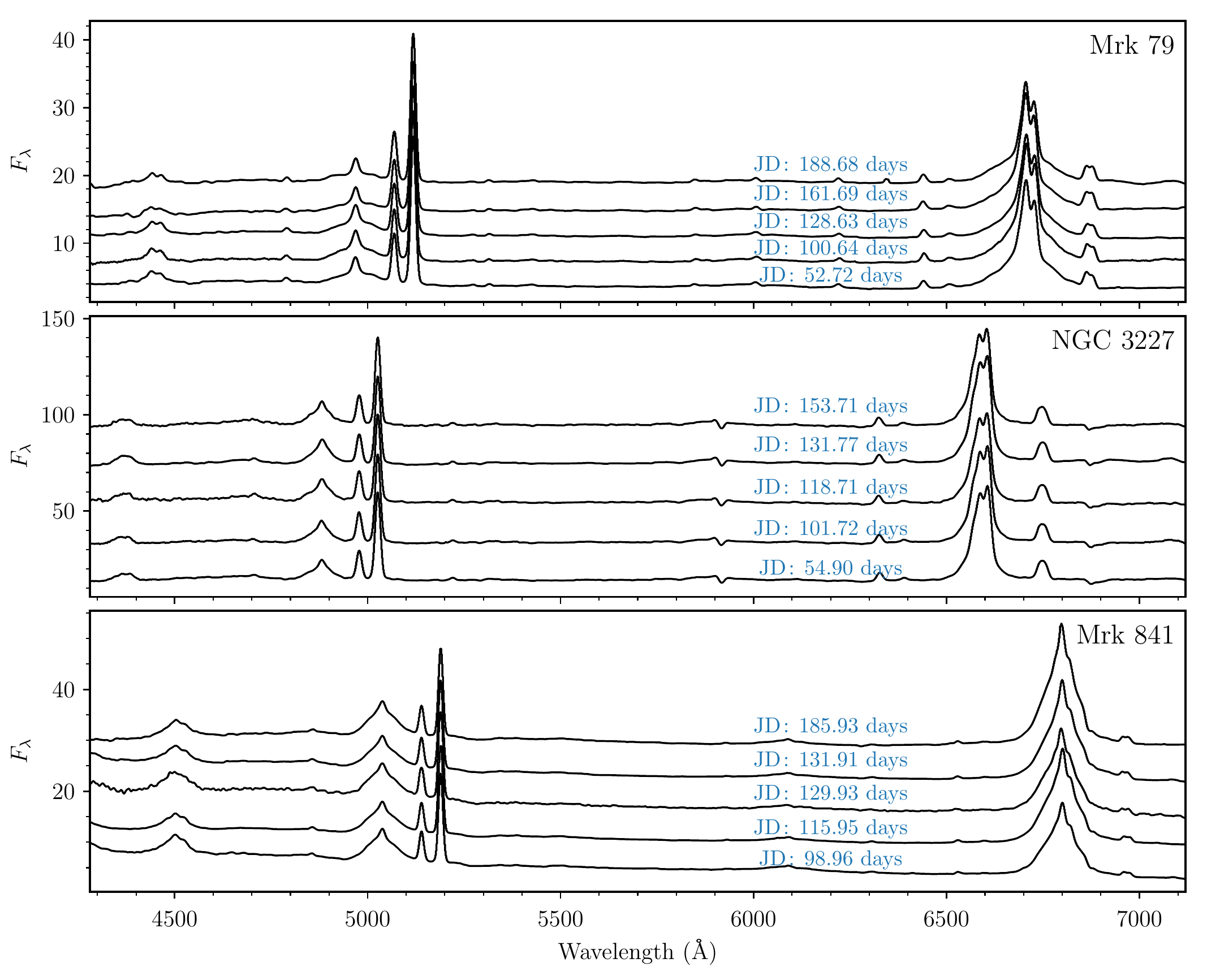}
    \caption{Some examples of the flux-calibrated spectra. The spectra are randomly selected from the datasets for each object. 
    The unit of flux density is $10^{-15}\ \mathrm{erg\ s^{-1}\ cm^{-2}\ \AA^{-1}}$. The spectra are shifted a little for clarity.
    Julian dates are from 2,457,700.}
    \label{fig:spectra}
\end{figure*}

Here we present 5 spectra that are randomly selected for each of the 3 objects in Figure \ref{fig:spectra}.

\section{The reconstructed continuum light curves}
\label{app:recon_LC}

\begin{figure*}
    \centering
    \includegraphics[width=0.95\textwidth]{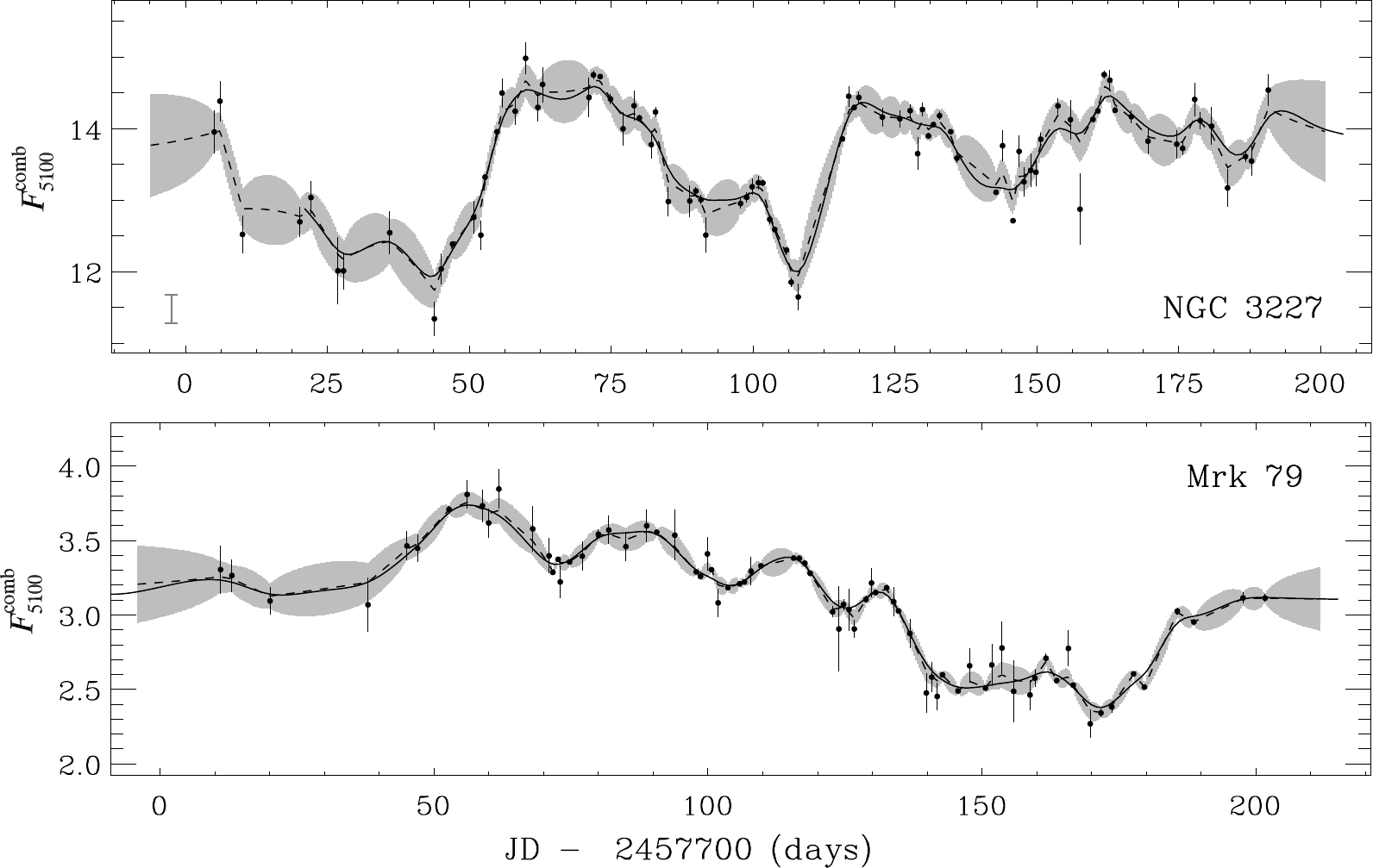}
    \caption{The input reconstructed continuum light curves for MEM. The points are the combined continuum light curves in Figure \ref{fig:light_curves}. The dashed lines and the grey shades are the reconstructed light curves and the corresponding uncertainties by DRW. The 
    solid lines are the MEM reconstructed results. The grey error bar is the systematic uncertainty in Figure \ref{fig:light_curves}.}
    \label{fig:recon_LC}
\end{figure*}

In Section \ref{sec:2dvdms}, we first used damped random walk to model the continuum, and adopted the reconstructed highly sampled continuum
light curves as the input for MEM. Here we present the reconstructed continuum light curves in Figure \ref{fig:recon_LC}.

\end{document}